\documentstyle[aasms4]{article}
 
\lefthead{Stout-Batalha and Batalha}
\righthead{Disk Accretion and Lithium Enhancements}
 
\begin{document}
 
\title{Accretion-Induced Lithium Line Enhancements in  \\ Classical T Tauri Stars: 
RW Aur}
 
\author{N.M. Stout-Batalha and C.C. Batalha}
\affil{Observat\'orio Nacional/CNPq, Rua General Jos\'e Cristino, 77, Rio de Janeiro, RJ 20920, Brasil \\
natalie@on.br, celso@on.br}
\author{G.S. Basri}
\affil{University of California, Berkeley, CA 94720 \\
basri@soleil.berkeley.edu}
 
\begin{abstract}

It is widely accepted that much of the stochastic variability of T Tauri stars
is due to accretion by a circumstellar disk.  The emission line spectrum as
well as the excess continuum emission are common probes of this process.  In
this communication, we present additional probes of the circumstellar environment
in the form of resonance lines of low ionization potential elements.
Using a set of 14 high resolution echelle observations of the classical T Tauri
star (CTTS), RW Aur, taken between 1986 and 1996, we carefully measure the 
continuum veiling at each
epoch by comparing more than 500 absorption lines with those of an appropriate
template.  This allows us to accurately subtract out the continuum emission
and to recover the underlying photospheric spectrum.  In doing so, we find
that selected photospheric lines are enhanced by the accretion process,
namely the resonance lines of LiI and KI.  A resonance line of TiI and a low 
excitation potential line of CaI also show weak enhancements. Simple slab
models and computed line bisectors lead us to propose that these line 
enhancements are markers of cool gas at the beginning of the accretion flow
which provides an additional source of line opacity.  These results
suggest that published values of surface lithium 
abundances of classical T Tauri stars are likely to be overestimated. This would
account for the various reports of surface lithium abundances in excess of 
meteoritic values among the extreme CTTS.  Computing LTE lithium abundances of
RW Aur in a low and then high accretion state yields abundances which vary 
by one order of magnitude.  The low accretion state lithium abundance is
consistent with theoretical predictions for a star of this age and mass while
the high accretion state spectrum yields a super-meteoritic lithium abundance.

\end{abstract}


\section{Introduction}

When an overabundance of lithium (a factor of 100 greater than in the Sun)
was first detected in a small sample of T Tauri stars, it was speculated
that either the circumstellar environment of these stars furnishes 
fresh lithium, or that the stars are too young to have undergone 
appreciable lithium destruction (\cite{bg60,z72}).
The latter argument is supported by the meteoritic lithium abundances 
which, at $\epsilon (N_{Li})=3.3$ (\cite{nm74}), are also found to be 
several orders of magnitude greater than that of the Sun at
$\epsilon (N_{Li})=1.16 \pm 0.1$ (\cite{ag89}).
Bodenheimer (1965) laid out the physics for the destruction of
lithium in late-type stars whereby surface lithium is carried
via convection to interior regions hotter than $\sim 2.4
\times 10^6$ K.  At such high temperatures, lithium is destroyed
in (p,$\alpha$) reactions.  The idea that surface lithium abundances
could be used as age indicators became widespread.  The question
became: exactly how and when does lithium depletion take place? 

Early theoretical work, such as that of Bodenheimer, predicted
that the surface lithium depletion would be a function of stellar
mass and age.  However, observations of coeval populations in young 
open clusters revealed that the relation would not be so simple.  
Duncan \& Jones (1983), in a study of F, G, and K-type stars in the 
Pleiades cluster ($\sim 100$ Myr), found a spread in the lithium abundance
(as much as 1.0 dex) for a given mass.  Since that time, extensive
studies of open cluster stars of varying ages and masses have 
emerged in the literature.  From studies of IC 2602 at 30 Myr (\cite{r96}), 
IC 4665 at $\sim 50$ Myr (\cite{mm97}), the $\alpha$ Persei cluster at
70 Myr (\cite{b88,r98,bm99}), the Pleiades at 120 Myr 
(\cite{s93a,g94,j96,bmg96,s98}), NGC 2516 at 150 Myr (\cite{jjt98}), 
NGC 6475 at 220 Myr (\cite{jj97}), NGC 1039 at 250 Myr (\cite{j97}), the Hyades
at 700 Myr (\cite{c84,t93,s95}), the Praesepe cluster
at 700 Myr (\cite{s93b}), and the much older cluster, NGC 2682, 
at $\sim 4$ Gyr (\cite{j99}), we have consistently seen that a
spread in lithium abundances at any one mass exists (for $M < 0.9 M_\odot$) 
up to the age of the Hyades.  We also find that the stars with the highest
rotation rates are all lithium-rich.

It has been speculated that rotation plays a key role in halting
lithium depletion at least up until the age of NGC 1039 (250 Myr).
Between this and the Hyades age, the ultra-fast rotating stars
(loosely defined as those with $v\sin i > 30$ km/s) will have spun-down
into the classical main-sequence distribution (\cite{k67}), 
and the corresponding surface lithium abundances will have settled
down into a relatively well-behaved distribution with little scatter
at any given mass.  Unfortunately, though, current theoretical models
have not successfully explained the, perhaps fortuitous, connection
between rotation and lithium depletion.  In fact, rotation actually 
seems to produce additional mixing thereby slightly enhancing the
depletion rate.  This problem has been well-summarized in the work
of Mendes, D'Antona \& Mazzitelli (1999).  Ventura et al. (1998) suggest
that the missing link is the stellar magnetic field, previously disregarded
in theoretical models.   A strong, rotationally induced, dynamo field could
inhibit convective mixing.  

Extending the observational studies to younger populations of stars --
the T Tauri stars (TTS) -- we look for additional clues.  We find that the
angular momentum distribution of these stars shows a systematic
difference between the mean rotation rate of the weak-lined T Tauri
stars (those lacking accretion from a circumstellar disk) and 
that of the classical T Tauri stars which are actively accreting disk
material (\cite{rv83,coy1,coy2,ch96}).  The weak-lined TTS (WTTS), 
on average, have shorter rotational periods than the classical TTS.
Edwards et al. (1993) hypothesize that disk accretion plays an integral role in 
regulating the angular momentum of the classical TTS as they contract
towards the main sequence.  The authors suggest that those TTS whose 
accretion is halted early on the Hayashi track would lack this effective 
brake and arrive at the zero-age main sequence as rapid rotators.  
Since all rapid rotators (with $M < 0.9 M_{\odot}$) at the ZAMS are 
lithium-rich compared to average depletion curves and since lithium
depletion occurs on the pre-main sequence (in the range $0.2-0.9 
M_{\odot}$; Mart\'{\i}n et al. 1994), we look for any systematic
differences in the lithium abundance patterns of WTTS versus CTTS.

Two studies comparing the distribution of lithium abundances of
classical and weak-lined T Tauri stars have been presented.   Basri, 
Mart\'{\i}n, \& Bertout (1991), in a study of 28 TTS in the Taurus-Auriga
star-forming complex, find no difference in the distribution of lithium 
abundances among classical and weak-lined TTS.  Magazz\`u, Rebolo, \& 
Pavlenko (1992) report similar results in their study of 36 TTS in the
Ophiucus, Taurus-Auriga, NGC 2264, Chamaleon, and Lupus star-forming
regions.  Interestingly, Basri et al. comments that the most extreme
classical TTS in their sample appear to be the most lithium-rich objects.
And looking more closely at the Magazz\`u et al. sample, the classical TTS
appear to be slightly more li-enhanced (on average) than the weak-lined TTS.
While the authors do not give significance to these results, they do point
out that the classical TTS, Sz 24, is extremely lithium-rich with an
abundance far in excess of even the meteoritic value of the Solar system.
They speculate that there might be an additional process at work
providing a fresh source of lithium.
Therefore, if there is any systematic difference between the classical
and weak-lined T Tauri stars, it is not evident from these studies, and, 
indeed, the opposite appears to be more likely.

The objective of this study is to look more carefully at the effects
of the accretion process on measured lithium abundances.  Studies
of lithium abundances among classical TTS are fraught with pitfalls.
In particular, the continuum veiling, produced by the accretion process, 
dilutes the stellar spectrum, making it difficult to reconstruct the 
true photospheric absorption.   We attempt, in this study, to accurately
measure the veiling in one classical TTS: RW Aur (K4; $\sim 1 M_\odot$).  
This star was chosen
since it is known to undergo extreme variations in its accretion as 
measured by the veiling.  We would expect that, if accurate veiling 
measurement are made and the appropriate corrections are made to the
observed spectrum, then reliable lithium abundances could be derived.
Fourteen echelle observations of RW Aur were collected at various epochs 
between 1986 and 1996.  Each observation presents a different veiling
level.  Though the accretion rate may be variable, we would expect to be
able to recover a consistent surface lithium abundance.  We do not.
The apparent abundances change in an unexpected manner which we
report herein.  We discuss the implications this has on lithium
depletion and accretion models.

\section{Observations}

High resolution echelle observations of RW Aur were obtained on 29 November
through 1 December 1996 using the Hamilton spectrograph (\cite{v87}) fed by the
3-meter Shane telescope of Lick Observatory.  Eighty-eight echelle orders
spanning a wavelength range from 3820 \AA~ to 9275 \AA~ were recorded on
a 2048 $\times$ 2048 thick Phosphor-coated CCD.   The observations were made using
a 640 $\micron$ slit width ($1\farcs2$ projected onto the sky) yielding $\sim 2$ pixel 
sampling (a resolving power of 56,000).  An observing log is given in Table~\ref{obs} 
which includes the time at mid-exposure (expressed as a heliocentric Julian date), 
the length of the exposure (in seconds), and the signal-to-noise per pixel at the
blaze peak.  These observations are supplemented by previously published
data of RW Aur which are also listed in Table~\ref{obs} and discussed in
more detail below (\S 2.2).

\begin{deluxetable}{cccccc}
\tablewidth{0pt}
\tablecaption{Observing Log of RW Aur \label{obs}}
\tablehead{
\colhead{HJD} &
\colhead{t$_{exp}$} &
\colhead{S/N} &
\colhead{v$_{rad}$} &
\colhead{v$\sin i$} &
\colhead{v \tablenotemark{a}} \\
\colhead{} & 
\colhead{[s]} &
\colhead{} &
\colhead{[km s$^{-1}$]} &
\colhead{[km s$^{-1}$]} &
\colhead{($\lambda \sim 6000$)}
}
\tablecolumns{5}
\startdata
2450417.80447	& 3600	& 60 & -2 & 22.7 & $2.7 \pm 1.0$  \nl
2450417.94248	& 3600	& 50 &  0 & 20.5 & $1.9 \pm 0.9$  \nl
2450419.73527	& 4500	& 60 & 22 & 23.3 & $0.6 \pm 0.4$  \nl
2450419.84735	& 4500	& 45 & 18 & 22.7 & $1.0 \pm 0.6$  \nl
2450420.02697	& 5400	& 80 & 17 & 23.5 & $1.8 \pm 0.8$  \nl
\cutinhead{previously published data \tablenotemark{b}}
21 Oct 1986  & \nodata & \nodata & \nodata & \nodata & $0.3 \pm 0.3$ \nl
21 Dec 1986  & \nodata & \nodata & \nodata & \nodata & $4.8 \pm 1.6$ \nl
07 Nov 1987  & \nodata & \nodata & \nodata & \nodata & $0.8 \pm 0.4$ \nl
04 Feb 1988  & \nodata & \nodata & \nodata & \nodata & $2.4 \pm 0.9$ \nl
30 Nov 1988  & \nodata & \nodata & \nodata & \nodata & $2.0 \pm 0.9$ \nl
17 Jan 1989  & \nodata & \nodata & \nodata & \nodata & $6.1 \pm 1.7$ \nl
17 Jan 1989  & \nodata & \nodata & \nodata & \nodata & $5.0 \pm 1.4$ \nl
17 Jan 1989  & \nodata & \nodata & \nodata & \nodata & $4.4 \pm 1.3$ \nl
14 Oct 1989  & \nodata & \nodata & \nodata & \nodata & $4.4 \pm 1.6$ \nl 
\enddata
\tablenotetext{a}{Throughout this communication, we refer to the veiling --
the ratio between the excess continuum emission and the photospheric continuum
emission -- using the symbol, v (lower case).  This is to avoid confusion with 
the apparent magnitude , represented by V (upper case).  We do not adopt the
more commonly used symbol, r, in order to avoid confusion with the Pearson's 
r-coefficients listed in Table~\ref{eqw}.}
\tablenotetext{b}{From \cite{bb90}, in which radial and rotational velocities
are computed for a single observation and used for all spectra.}
\end{deluxetable}

\subsection{Reduction and analysis}

Data are reduced using the Image Reduction and Analysis Facility (IRAF).
Thermal dark current corrections are found to be unnecessary. Pixel-to-pixel
gain variations are removed using several spectra (combined and normalized)
 of a quartz-halogen incandescent lamp, filtered for more evenly distributed 
signal at all wavelengths.  Pixel-to-pixel gain variations are typically less 
than $\pm 3\%$ in amplitude.  Echelle orders are extracted following the optimal 
extraction algorithm of Horne (1986) and Marsh (1989).  Wavelength
calibration is performed by identifying over 3000 lines in the spectrum
of a Thorium-Argon lamp taken with the same instrumental setup.  A two
dimensional polynomial fit gives the dispersion solution whose residuals
(RMS) are less than \hbox{$\,^1\!/_{10}$} of a pixel.

Spectral standards, taken from the catalog of Keenan \& McNeil (1989),  were 
also observed at this epoch to aid in the spectral classification of the
program star as well as in the determination of radial velocity.  Table~\ref{std}
lists these standards and their relevant parameters.  The featureless spectrum 
of a rapidly rotating B star, $\epsilon$~Per, was observed for the purpose of 
removing telluric lines from the spectra of the program star as deemed necessary.
This is found to be particularly relevant for the potassium line 
(KI~$\lambda~7698.977$~\AA).  The spectral order containing this wavelength region
is normalized, shifted to zero, and then scaled iteratively until the telluric
lines are optimally removed from the spectrum of the program star upon
subtraction.

\begin{deluxetable}{llcccccc}
\tablewidth{0pt}
\tablecaption{Spectral Standards \label{std}}
\tablecolumns{7}
\tablehead{
\colhead{} & \colhead{} & \colhead{} & \colhead{} & \colhead{} & \colhead{Catalog} & \colhead{This work} \\
\colhead{HD} &
\colhead{GJ} & 
\colhead{$\alpha$(2000)} &
\colhead{$\delta$(2000)} &
\colhead{SpTyp \tablenotemark{a}} &
\colhead{v$_{rad}$ \tablenotemark{b}} &
\colhead{v$_{rad}$} &
\colhead{v$\sin$ i \tablenotemark{c}} \\
\colhead{} &
\colhead{} &
\colhead{} &
\colhead{} &
\colhead{} &
\colhead{[km s$^{-1}$]} &
\colhead{[km s$^{-1}$]} &
\colhead{[km s$^{-1}$]}
}
\startdata
\cutinhead{November 1996}
206301	& \nodata & 21~41~33.19 & -14~02~36.1 & G2 IV & -1.2 \tablenotemark{e} & -10.2 & \nodata \nl
191026	& \nodata & 20~06~22.59 & +35~58~43.0 & G8.5 IV & -33.4 \tablenotemark{e} & -32.9 & \nodata \nl
198809	& NN 4168 & 20~52~07.84 & +27~05~52.2 & G7 III & +1.0 & +8.1 & \nodata \nl
117176	& 512.1 & 13~28~26.54 	& +13~47~12.4 & G4 V   & +5.3 & +5.2 & \nodata \nl
101501	& 434  	& 11~41~02.99 	& +34~12~24.8 & G8 V   & -5.9 & -5.3 & \nodata \nl
188376	& \nodata & 19~55~49.51 & -26~18~02.2 & G5 IV  & -15.4 \tablenotemark{e} & -14.9 & \nodata \nl	
10700	& 71   	& 01~44~09.99 	& -15~56~57.9 & G8 V   & -17.0 & -16.7 & \nodata \nl
3651	& 27   	& 00~39~23.36 	& +21~15~20.1 & K0 V   & -32.9 & -31.4 & 0.0 \nl
10476	& 68   	& 01~42~30.77 	& +20~16~40.3 & K1 V   & -33.7 & -33.3 & \nodata \nl
20630	& 137  	& 03~19~20.73 	& +03~22~07.9 & G5 V   & +18.9 & +19.2 & \nodata \nl
\cutinhead{October 1990/January 1991}
3765	& 28	& 00 40 47.61	& +40 11 46.9 &	K2 V   & -64.2 & \nodata & 0.0 \nl
223778	& 909A	& 23 52 20.85	& +75 32 38.2 & K3 V   & +1.7  & \nodata & \nodata \nl
88230	& 380	& 10 11 29.04	& +49 27 40.6 & K6 V   & -25.4 & \nodata & \nodata \nl
\nodata	& 908.1	& 23 50 26.79	& +30 21 10.9 & K4 V \tablenotemark{d} & -3.1  & \nodata & 3.6 \nl 
19305	& 123	& 03 06 25.37	& +01 58 41.6 & M0 V   & -30.2 & \nodata & \nodata \nl 
36395	& 205	& 05 31 24.76	& -03 38 54.0 & M1.5 V & +8.5  & \nodata & \nodata \nl
95735	& 411	& 11 03 22.52	& +36 02 10.2 & M2 V   & -84.3 & \nodata & \nodata \nl
\cutinhead{Miscellaneous}
16160	& 105A	& 02 36 04.89 	& +06 53 12.7 & K3V	& +26.0 & \nodata & 2.0 \nl
32147	& 183	& 05 00 49.00 	& -05 45 13.2 & K3V \tablenotemark{e} & +22.2 & \nodata & 0.0 \nl 
88230	& 380	& 10 11 22.14	& +49 27 15.3 & K6V	& -25.4 & \nodata & 1.9 \nl
131977	& 570A	& 14 57 28.00	& -21 24 55.7 & K4V	& +29.1 & \nodata & \nodata \nl
201091	& 820A	& 21 06 53.94	& +38 44 57.9 & K5V	& -64.8 & \nodata & 1.3 \nl
201092	& 820B	& 21 06 55.26	& +38 44 31.4 & K7V	& -64.3 & \nodata & 1.1 \nl
208527	& 841.1 & 21 56 23.98	& +21 14 23.5 & K5V \tablenotemark{e} & +2.0 \tablenotemark{e} & \nodata & \nodata 
\enddata
\tablenotetext{a}{all but those marked taken from \cite{km89}} 
\tablenotetext{b}{all but those marked taken from \cite{gliese}}
\tablenotetext{c}{taken from \cite{f98}}
\tablenotetext{d}{\cite{f98}}
\tablenotetext{d}{from SIMBAD database}
\end{deluxetable}

\subsection{Previously published data}

Also included in this analysis are high resolution echelle spectra of RW Aur
and various spectral standards taken
between 1986 and 1991 by Gibor Basri and collaborators also 
with the 3-meter Shane telescope of Lick Observatory and the Hamilton 
spectrograph.  The description of the RW Aur observations can be found in
Basri \& Batalha (1990) and Batalha et al. (1996). The additional spectral standards are
listed in Table~\ref{std}.  It is worth noting a few significant 
differences between these data and the more recent observations taken November
1996.  In 1994, the imaging optics of the Hamilton spectrograph were 
refurbished resulting in a narrower and more symmetric instrumental profile
as well as a larger separation between echelle orders allowing for a more
accurate determination of the scattered light surface. 
In addition, the older RW Aur observations were recorded on a 
Texas Instruments 800 $\times$ 800 CCD resulting in 
a more restricted spectral coverage as well as discontinuous wavelength 
coverage from one echelle order to the next. 

\subsection{Radial and Rotational Velocities}

The radial velocity of RW Aur is determined for each independent observation
of the 1996 epoch
using a cross-correlation method similar to that outlined by Tonry \& Davis
(1979).  The spectral standard, HD 3651 (K0, v$_{rad} = -32.9$), is used as
the template.  Its radial velocity is confirmed using the solar spectrum
as reference.  All echelle
orders are used simultaneously in computing the cross-correlation function
after filtering out atmospheric bands and stellar emission lines.
The peaks of the cross-correlation functions define the radial velocities.
Radial velocities are also computed for the spectral
standards observed during the 1996 epoch to confirm the accuracy of the
method.   Both the catalog values and our determinations are listed in
Table~\ref{std}.  For all but two stars, the differences are less than
1 km s$^{-1}$.  Both HD 206301 and HD 191026 are classified as RS CVn stars,
and are therefore expected to show radial velocity variations.  And previously
published radial velocity measurements of HD 198809 show a peak-to-peak variation
of $\sim 10$ km s$^{-1}$.

Radial velocities of the RW Aur observations are listed in Table~\ref{obs}.
The associated cross-correlation functions for the Nov 1996 observations
are plotted in Figure~\ref{xc}.  We find peak-to-peak variations of more
than 20 km s$^{-1}$ -- a result also obtained by Hartmann et al. (1986) --
suggesting the possibility of a close companion.  RW Aur is already known
to be a multiple system with A and B separated by approximately $1\farcs4$
and B and C separated by approximately $0\farcs12$.  With a flux ratio in the
optical of 68:1 (\cite{gws97}), we do not expect to find contamination from 
RW Aur B in the spectrum of RW Aur A.  Ghez, White, \& Simon (1997) have not
found evidence of a closer companion brighter than K=10.6 down to $0\farcs1$.
However, assuming a distance of 140 pc to RW Aur, a resolution of 
$0\farcs1$ corresponds to a separation of more than 6 AU.  Since we see 
variations of more than 20 km/s over a period of just two days, we expect
a closer companion to be the cause.  Since there is no
evidence of a secondary set of lines, we proceed with the assumption that
the observed spectra are not contaminated by such a companion. 

The projected rotational velocities are determined for the 1996 observations 
after spectral type 
determination (see \S 3).  The spectrum of GJ 380 is used as the template.
The entire spectrum is artificially broadened in
steps of 5 km s$^{-1}$ and cross-correlated with the original spectrum.
The FWHM of the resulting cross-correlation functions are then measured
as a function of rotational velocity.  RW Aur spectra are then 
cross-correlated with the template and the FWHM measured.  This FWHM
is then interpolated along the previously defined distribution thereby
defining the projected rotational velocity.  The entire echelle spectrum is 
strung together to make one vector which is used to compute the 
cross-correlation function.  This results in a very high quality
function with little noise.  However, in practice, it is quite difficult
to properly filter out all of the stellar emission lines, especially for
those observations of RW Aur at high accretion rates.  Emission lines
produce asymmetries in the cross-correlation functions which produce
systematic errors in the $v\sin i$ determinations.  It is therefore
plausible that the dispersion in the measurements listed in Table~\ref{obs}
are due to contaminating emission lines.  This is currently being investigated
further.

\begin{figure}
\caption{Normalized cross-correlation functions for the five observations taken
in November 1996, plotted with the following symbols: HJD 2450417.80447 - 
diamonds; HJD 2450417.94248 - triangles; HJD 2450419.73527 - solid line;
HJD 2450419.84735 - asterisks; HJD 2450420.02697 - crosses.}
\label{xc}
\end{figure}

\section{Spectral Type Estimation}

For an accurate analysis of the continuum veiling, it is necessary
to deduce the spectral type of RW Aur to within 2 subclasses.
Errors larger than this can lead to systematic veiling errors as low (versus high)
excitation potential lines yield consistently low (or high, depending
on the spectral type in question) results. They also work to increase the overall
scatter in veiling values measured using large numbers of 
absorption lines with various dependences on temperature.

To estimate the spectral type of RW Aur, we construct a grid of
spectral standards ranging from G2 to M2 (refer to Table~\ref{std}).  
We select the observation
of RW Aur which shows, upon simple inspection, the least veiled spectrum
(HJD=2450419.73527).  
Spectral orders free of telluric lines and stellar emission lines are 
pre-selected.  A result will be obtained for each of these echelle orders
which will then be averaged together with all other results for the final 
solution.

We expect that even this low-veiled observation of RW Aur will have 
some excess continuum emission contaminating the underlying spectrum.
We therefore proceed as follows.  For each of the pre-selected echelle
orders, all spectral standards are artificially veiled until the best
match (lowest $\chi ^2$ statistic) to the corresponding wavelength window 
of RW Aur is obtained.
The normalized $\chi ^2$ is computed for each of these solutions and that
which produces the best fit is noted.  In this manner, we obtain an initial
veiling estimate and a spectral type for each of the spectral orders 
we have selected, 59 in all.  GJ 908.1 provides the best match to the
spectrum of RW Aur, with a veiling estimate of $0.8\pm 0.5$. 

GJ 908.1 is classified as K8V in the catalog of Gliese \& Jahreiss (1991). 
Unfortunately, this is the only star in our sample
of standards that is not classified in the Perkins Catalog of Revised 
MK Types (\cite{km89}).  However, in a recent, detailed analysis of 
K-dwarfs in the Solar neighborhood, Fischer (1998) computes (via spectral
synthesis techniques) an effective temperature for this object of 4556 K --
more than 400 K larger than the color temperature listed in the Gliese
\& Jahreiss catalog.  This temperature (corresponding to K4) is confirmed 
using line ratio techniques. 
We therefore adopt a spectral type of K4 for RW Aur, consistent with 
the independent determination of Basri \& Batalha (1990).

\section{Veiling Determinations}

\subsection{The Method}

The method used herein for determining precise veiling values is loosely based on the
method described by Basri \& Batalha (1990).  A catalog of more than
1000 lines was constructed, and the veiling is computed on a line-by-line
basis using GJ 908.1 (\S 3.0) as the unveiled reference.
After making minor adjustments to the continuum
normalization (relative to a local continuum), the template line
is convolved with a rotational kernel to match the projected rotational
velocity of the program star (20 km s$^{-1}$).  Gaussian fits are then
made to the program and template lines in question.  The template gaussian
is artificially veiled until a $\chi ^2$ minimum is located defining
the veiling value for that line.  

\subsection{Veiling Results}

Figure~\ref{veiling} displays the resulting veiling values as a function
of the rest wavelength of the line.  Also shown are linear fits
to the distributions.  There is a general trend for the veiling to 
increase slightly toward bluer wavelengths.  However, a single veiling value 
for each observation is computed by taking an average of all points between
5500 \AA~ and 6500 \AA.  These values are tabulated in Table~\ref{obs} and listed in 
Figure~\ref{veiling}. Also
listed are the number of lines yielding measurements for each observation.
Obviously, as the continuum of RW Aur becomes more heavily veiled, the
number of absorption lines we can detect decreases substantially (from 
more than 500 down to merely 35 in this case) and 
the noise becomes more significant.  Furthermore, the older Hamilton data
do not contain as broad of spectral coverage as do the newer observations.
All of these factors are reflected in the dispersion (sample standard 
deviation) measurements which we use to estimate the uncertainty.  

\begin{figure}
\caption{Veiling values as a function of wavelength.  Each symbol represents
the veiling value determined for one particular absorption line.  Polynomial
fits (1st order) are made to the data and the standard deviation relative to 
each fit is computed and listed as the uncertainty.  The veiling value listed
above each plot is the mean value between 5500 \AA~ and 6500 \AA.}
\label{veiling}
\end{figure}

Veiling determinations for the older data have been previously published
(\cite{bsbt96}).  Our results are consistent to within the determined 
uncertainties.

\subsection{Line-to-line trends}

We check for any systematic line-to-line trends that may suggest the
use of an inappropriate template.  A difference in magnetic activity levels, 
for instance, may modify the T-$\tau$ relation in the TTS relative to
the inactive template resulting in generally shallower line profiles.
Lines formed closer to the temperature minimum will suffer more than those
formed deeper in the photosphere yielding systematic errors in veiling
(\cite{fb87,bws89}).  
Errors in effective temperature also yield systematic errors due to the fact
that different lines have different dependences on temperature (\cite{hg97}).  

Assuming that we have a large sample of lines and given the fact that
we are more  concerned with relative as opposed to absolute veiling
values, neither of these 
errors should compromise the science objectives of this project.
However, this is clearly not the case as we get to very large veiling
values where the number of lines employed decreases significantly.
Furthermore, as we look at the more heavily veiled observations, 
we find we are restricted to the strongest of absorption lines which 
may further conspire to produce systematic errors.

Plotting all of the data for a low veiling case, we
find that the veiling tends to increase towards bluer
wavelengths as we would expect if the continuum source
is hotter that the stellar photosphere (Figure~\ref{test} a).
We also note a slight increase in the average excitation potential 
at redder wavelengths (Figure~\ref{test} b).  This casts some doubt
on the wavelength dependence of the veiling distribution
since our program star is undoubtedly more magnetically active
than the template star.  It is therefore possible that lines of
low excitation potential yield systematically higher veiling
values than lines of high excitation potential thereby
driving the veiling distribution down towards redder wavelengths.
 
To check this, we quantify the veiling distribution considering only
low and then only high excitation potential lines.  Polynomial fits to
lines of $\chi _0 > 4.5$ (56 lines) and $\chi _0 < 0.9$ (51 lines)
are computed and displayed in Figure~\ref{test} c.  A statistically significant
increase towards the blue is found in both cases while the 
magnitude of the veiling decreases slightly (by $\sim 0.22$) for the 
high excitation
potential sample.  We conclude that the increase of the veiling
distribution towards the blue does reflect the temperature of the
emitting source.  We also suspect that the absolute veiling value
is influenced by a mis-match in the magnetic activity levels of the
program star and the template, though the effect is smaller than the
intrinsic (rms) error in the veiling measurement (0.4).  Since, in this
study, we are concerned with veiling variations significantly larger
than 0.4, we conclude that this effect does not compromise our results
even in the extreme case that our measurements for the high veiling
observations are restricted only to low excitation potential lines.
This, however, does not turn out to be the case.  The filled symbols in Figure~\ref{test} b  mark
the lines used in the highest veiling observation (17 Jan 89) of RW Aur.
We find we actually use more high excitation potential lines to measure the
veiling of this observation.
 
To quantify possible errors in effective temperature, we compute the
veiling distribution for the HJD 2450419.73527 observations ($v=0.6\pm 0.4$) using a
K2 template and then an M2 template.  Redward of 5500 \AA, the
K2 template gives an average veiling value of $0.5\pm 0.3$ while the
M2 yields $0.3\pm 0.4$.  As expected, the measured veiling value increases slightly
going from the K2 to K4 template reflecting the fact that most lines are
getting deeper as the star gets cooler.  As we look toward an even
cooler star (M2), a significant number of lines 
become shallower once more and the average veiling value
decreases accordingly.  Again, the errors
we find arising from improper spectral classification are smaller
than the intrinsic scatter of the measurements and significantly smaller
than the range of variations RW Aur undergoes.

\begin{figure}
\caption{{\bf a)} Veiling measurements of HJD 2450419.73527 (v=0.6) taken
from Figure~\ref{veiling}, {\bf b)} Excitation potential (low) of lines 
represented in a) as a function of wavelength.  The linear fit shows
the slight increase of the average EP towards redder wavelengths.  Filled
circles mark the lines used to quantify the high-veiling observations of 
17 Jan 89, {\bf c)} The lowest EP lines ($\chi _0 < 0.9$)
from a) are plotted (open circles) together with the corresponding linear
fit (dashed line).  The highest EP lines ($\chi _0 > 4.5$) are also plotted
(filled circles) together with their linear fit (solid line).  Both distributions
show an increase in veiling towards the blue.  The high EP lines present slightly
lower veiling values, though the difference is smaller than their intrinsic
(rms) dispersion.}
\label{test}
\end{figure}
 
\section{The Photospheric Lines}

We select a set of photospheric absorption lines which sample 
a relatively wide range of excitation potentials and which
are not severely blended within the rotationally broadened 
line width.  These lines are listed in Table~\ref{eqw}. Weak
(but significant) subordinate lines which lie within the 
rotational Doppler width are listed in parentheses.
 
The line equivalent width is measured for each observation
of RW Aur.  In order to de-veil the spectrum and recover
an estimate of the true underlying photospheric line
strength, we multiply each equivalent width by a factor
of (1+v), where {\it v} is the veiling listed in Table 1.  As shown 
in Basri \& Batalha (1990), this is
the line strength one would obtain after subtracting off
the excess continuum emission distribution.  We make a
simplification in these calculations which is to assume one average 
veiling value (given in Table~\ref{obs}) for all lines redward 
of 5500 \AA.  Changes in veiling in this wavelength range are small.
The mean values of the de-veiled equivalent widths are listed in columns 4 
and 11 of Table~\ref{eqw} for each line.  In some cases, a line is not 
included in the spectral format of a particular observation (true for 
the older observations which were taken with a significantly
smaller CCD) or is contaminated by bad columns, etc.  Columns 6 and 13 list 
the number of observations for each line which yielded a measurement.  
Columns 3 and 10 list the (lower) excitation potential of the transition.

\begin{deluxetable}{crrcrrr|crrcrrr}
\tablewidth{0pt}
\tablecaption{Line Sample and Correlation Coefficients\label{eqw}}
\tablehead{
\colhead{ID} &
\colhead{$\lambda$} &
\colhead{EP} &
\colhead{EW \tablenotemark{a}} &
\colhead{{\it r} \tablenotemark{b}} &
\colhead{N \tablenotemark{c}} &
\colhead{{\it t} \tablenotemark{d}} &
\colhead{ID} &
\colhead{$\lambda$} &
\colhead{$\chi _0$} &
\colhead{EW \tablenotemark{a}} &
\colhead{{\it r} \tablenotemark{b}} &
\colhead{N \tablenotemark{c}} &
\colhead{{\it t} \tablenotemark{d}} \\
\colhead{} &
\colhead{\AA} &
\colhead{eV} &
\colhead{\AA} &
\colhead{} &
\colhead{} &
\colhead{} &
\colhead{} &
\colhead{\AA} &
\colhead{eV} &
\colhead{\AA} &
\colhead{} &
\colhead{} &
\colhead{}
}
\startdata
 CaI & 5598.48 & 2.52 & 0.34 & -0.166 & 14 & -0.584 & FeI & 7511.02 & 4.18 & 0.24 &  0.530 & 14 &  2.163 \nl
(FeI)& 5598.30 & 4.65 &      &        &    &        & FeI & 7937.13 & 4.31 & 0.22 &  0.170 & 14 &  0.596 \nl
 CaI & 6122.22 & 1.89 & 0.79 &  0.734 & 12 &  3.420 &  KI & 7698.97 & 0.00 & 1.93 &  0.964 & 14 & 12.618 \nl
 CaI & 6439.08 & 2.53 & 0.47 &  0.269 &  5 &  0.483 & LiI & 6707.76 & 0.00 & 0.95 &  0.902 & 14 &  7.224 \nl
 CaI & 6572.78 & 0.00 & 0.07 & -0.144 & 14 & -0.504 & MgI & 5528.41 & 4.35 & 0.14 & -0.482 & 14 & -1.904 \nl
(CrI)& 6572.89 & 1.00 &      &        &    &        & MnI & 5394.64 & 0.00 & 0.15 &  0.118 & 14 &  0.412 \nl
 CrI & 7400.23 & 2.90 & 0.22 & -0.107 & 14 & -0.372 &(FeI)& 5394.68 & 4.19 &      &        &    &        \nl
 FeI & 5762.99 & 4.21 & 0.22 & -0.032 & 14 & -0.111 & MnI & 6021.82 & 3.07 & 0.23 &  0.293 & 13 &  1.015 \nl
 FeI & 5934.66 & 3.93 & 0.14 &  0.145 & 14 &  0.509 &(FeI)& 6021.79 & 2.20 &      &        &    &        \nl
 FeI & 6003.01 & 3.88 & 0.21 &  0.260 & 14 &  0.933 & NaI & 5688.21 & 2.10 & 0.20 &  0.609 &  6 &  1.537 \nl
(VI) & 6002.62 & 1.05 &      &        &    &        &(NaI)& 5688.19 & 2.10 &      &        &    &        \nl
(TiI)& 6002.63 & 2.16 &      &        &    &        & NaI & 6154.23 & 2.10 & 0.09 &  0.319 & 13 &  1.116 \nl
 FeI & 6020.17 & 4.61 & 0.20 & -0.414 & 14 & -1.573 & NiI & 6643.63 & 1.68 & 0.09 & -0.212 & 14 & -0.752 \nl
 FeI & 6024.06 & 4.55 & 0.20 &  0.412 & 13 &  1.500 & NiI & 7122.19 & 3.54 & 0.12 & -0.452 & 14 & -1.755 \nl
 FeI & 6400.00 & 0.91 & 0.28 & -0.272 & 14 & -0.978 & TiI & 4997.10 & 0.00 & 0.39 &  0.693 & 14 &  3.332 \nl
(FeI)& 6400.32 & 0.92 &      &        &    &        &(NiI)& 4996.84 & 3.64 &      &        &    &        \nl
 FeI & 6430.85 & 2.18 & 0.15 & -0.180 & 13 & -0.607 & TiI & 5953.16 & 1.89 & 0.23 &  0.496 & 14 &  1.976 \nl
 FeI & 6633.75 & 4.56 & 0.09 &  0.508 & 14 &  2.042 &(FeI)& 5952.72 & 3.98 &      &        &    &        \nl
 FeI & 7130.92 & 4.22 & 0.15 &  0.277 & 14 &  1.000 & TiI & 6261.10 & 1.43 & 0.18 &  0.726 & 12 &  3.334 \nl
 FeI & 7495.06 & 4.22 & 0.17 &  0.014 & 14 &  0.048 & (VI)& 6261.23 & 0.27 &      &        &    &        \nl 
(FeI)& 7494.72 & 1.56 &      &        &    &        &     &         &      &      &        &    &        \nl
\enddata
\tablenotetext{a}{Average line strength over timeseries}
\tablenotetext{b}{Linear correlation coefficient, or Pearson's r-coefficient.}
\tablenotetext{c}{Number of measurements}
\tablenotetext{d}{Student's t-distribution}
\end{deluxetable}

Figure~\ref{allpts} plots all of the de-veiled EW's (not the mean values listed
in Table~\ref{eqw}) as a function of veiling.  For display purposes, we shift
each distribution vertically so that the linear fits to the distributions 
are zero at the lowest veiling value.  A large majority of the lines (24 out of 29, or $83\%$)
present a constant equivalent width value (within our uncertainty levels),
independent of veiling, as shown in the lower panel of Figure~\ref{allpts}.
This is what we'd expect to obtain assuming we have accurately recovered the purely
photospheric absorption line spectrum.  What is puzzling, however, is the behavior
of the five lines plotted in the upper panel of the same figure: KI $\lambda$ 7699.9
\AA, LiI $\lambda$ 6707.7 \AA, CaI $\lambda$ 6122 \AA, TiI $\lambda$ 4997 \AA, and
TiI $\lambda$ 6261 \AA.  Each of these lines appears to be enhanced by the accretion
process.  Figure~\ref{some} contains plots of the weaker four correlations, amplifying
the y-axes for better visualization. 
 
Linear correlation coefficients (Pearson's r-coefficients) are computed for each
set of veiling/de-veiled EW parameters.  These coefficients are listed in columns 5
and 12 of Table~\ref{eqw}.  To test the significance of each correlation, we compute the
Student's t-statistic (tabulated in columns 7 and 14).  Only the 5 lines mentioned
previously show a correlation at a confidence level better than $\alpha = 0.005$
(99.5\% probability that a correlation exists) as tested by this statistic.    
Two FeI lines ($\lambda 6633.8$ and $\lambda 7511$ \AA) show a correlation at the 
$\alpha = 0.05$ confidence level.  However, due to the fact that the errors are correlated
--that is, an error in veiling yields a systematic and predictable error in the deveiled
equivalent width -- we are required to demand higher significance levels.  Monte Carlo
simulations reveal the nature of the correlated errors, the results of which are shown in 
Figure~\ref{mc}.  The arrows mark the one-sigma deviations due to normally distributed
error estimates in both parameters.  Eight thousand randomly generated test points
are used to test the correlated errors.  The TiI $\lambda 6261$ line shows only a marginal
correlation at the 1-sigma level.  LiI $\lambda 6707$ shows a correlation at the 2-sigma
level while KI $\lambda 7699$ shows a correlation at the 3-sigma level (not shown in plot).  
And the remaining two lines, TiI $\lambda 4997$ and CaI $\lambda 6122$ show correlations at 
the 1-sigma level.

Given the large number of lines which show no correlation, and given the 
statistical tests described above, we conclude that the resonance lines
of KI, LiI, and TiI as well as the low excitation potential line of CaI
are indeed enhanced by the accretion process.  We emphasize not only the
low excitation potential but also the low ionization potential of these
elements (4.34 - 6.11 eV).   None of the low excitation potential lines 
of the higher IP elements, FeI, NiI, and MnI (7.43 - 7.87 eV), show any 
indication of a correlation with veiling at our detection level. Using the 
catalogue of solar lines (\cite{solcat}) and the VALD database, we search for other resonance 
lines which could contribute to our sample, especially those of low IP elements 
such as TiI, VI, RbI, and CsI.  Unfortunately, we are unable to identify any additional 
lines whose equivalent widths can be recovered at all (or nearly all) veiling 
values and whose rotationally broadened line width is not overwhelmingly blended.
We measure a weak resonance line of RbI ($\lambda 7800.3$ \AA) in the low
veiling observations.  Although this line is of great interest due to its
low ionization potential (4.18 eV), we are unable to detect measurable absorption 
in the highly veiled spectra.

\begin{figure}
\caption{Deveiled equivalent widths are measured for each line and each
observation of RW Aur and then plotted against the veiling.  24 out of
29 lines show no correlation between these two parameters (lower panel).
However, the 5 lines plotted in the upper panel all yield a significant
correlation as tested by computing linear correlation coefficients and
comparing these against the Student's t-distribution, both of which are
tabulated in Table~\ref{eqw}.}
\label{allpts}
\end{figure}

\begin{figure}
\caption{The upper panel of Figure~\ref{allpts} is reproduced here, once for
each of the LiI, CaI, and TiI lines, with the y-axis amplified to better 
illustrate these weaker correlations.} 
\label{some}
\end{figure}

\begin{figure}
\caption{Monte Carlo simulations are run to test the effect of correlated
errors in veiling (x-axis) and the de-veiled EW's (y-axis).  Arrows mark
the $\pm 1\sigma$ deviations.}
\label{mc}
\end{figure}

\section{Discussion}

\subsection{Line Enhancement Mechanisms}

We note that the lines showing accretion-induced enhancements are resonance lines
(or low excitation potential lines) of low ionization potential elements.  With this
in mind, we consider the possible mechanisms which may be at work.

Dark spots arising from localized concentrations of surface magnetic flux can
increase the disk-integrated equivalent width of temperature sensitive lines
(those who increase in strength going from photospheric to cooler spot 
temperatures).  This is the case for the resonance lines of lithium and
potassium.  However, the effect is predicted to be small (e.g. $\sim 7\%$
for KI $\lambda 7699$ \AA~ in a K6 photosphere with a 14\% spot filling factor, 
and 3000K spot temperature (\cite{mythesis})) 
compared to the large enhancements observed herein (300\% for lithium, for example).  
Observations also reveal that the disk-integrated effect of surface spots
is small.  Less than a 15\% change in line strength is observed in the 
lithium resonance line as the large area spots of HII 686 and HII 3163, two
ultra-fast rotating ZAMS stars of the Pleiades, rotate in and out of view
(\cite{sv99}).  Similar results are found for the rapidly rotating, weak-lined
T Tauri stars, P1724 (\cite{n98}) and Oph 052 and Oph 120S (\cite{p93}).  Considering the typically slow rotation rates of 
classical TTS (and RW Aur as implied by its small projected rotational velocity), 
we find it very unlikely that the line enhancements are a consequence of dark
spots.

We entertain the idea that the line enhancements are true elemental
abundance variations, perhaps furnished by disk accretion.  This is
perhaps appealing for the lithium enhancements since the surface
abundances may be slightly depleted through convective mixing for a
star of this estimated mass ($\sim 1 M_\odot$; \cite{bbb88}) and age 
(\cite{m99}).  However, this quickly
proves to be unlikely.  If we compute surface lithium abundances
using de-veiled lines (see \S 6.4), we recover values (at intermediate 
accretion rates) more than an order of magnitude larger than the primordial 
(meteoritic) lithium abundance.  Moreover, variations occur on timescales shorter
than one day.  We could not explain this considering the accretion
of a relatively small amount ($10^{-7}-10^{-8} M_\odot yr^{-1}$; \cite{v93,heg95}) of disk gas 
of no more than meteoritic number densities of lithium, 
onto the stellar surface.  Finally, we are still left with the additional problem of explaining 
the even larger potassium enhancements, impossible to reconcile if we are 
correct in assuming that the disk and stellar abundances of potassium should 
be nearly identical.  This should also be true for titanium.

We consider the presence of a low temperature component to the accretion
flow which could furnish a significant number of neutral species of the
elements in question.  
The gas temperature and density should yield
the strongest enhancements in potassium, followed by lithium and then 
titanium, with no enhancements measured in the higher IP element, manganese.
In the following section, we present simple slab models which suggest
that the presence of such a cool gas could explain the various enhancements
we observe.  We then compute line bisectors which suggest that the excess
absorption indeed arises in an infalling gas.

\subsection{Slab Models}

The distribution of cold gas around RW Aur is approximated by an
isothermal slab model characterized by temperature, density and 
physical depth. Line and continuum opacities are computed over a range of 
temperatures (1250 to 4000 K) and densities ($\log N_H = 11.0 - 17.0$ cm$^{-3}$)
while holding the physical depth (0.1 R$_*$) and surface area
(100\% of stellar surface) constant. The objective is to determine, qualitatively, if
a cold gas can yield the relative enhancements in the resonance lines of KI, LiI, 
TiI, and MnI.  We also include RbI $\lambda 7800.3$ \AA~ in the analysis even 
though we do not have a complete set of measurements (see \S 5) for this 
resonance line.  It is weak ($\sim 100$ m\AA) in the spectrum of our
K4 templates but becomes pronounced in the spectra of L stars (below
2200 K).  The fact that we do not detect this line in the highly veiled observations
is an additional constraint on the gas temperature in the slab.
Standard continuum opacity sources are included as a means of checking
the global stellar extinction.  We demand that the circumstellar gas remains
optically thin in the continuum.  Level populations which control the line 
opacities are computed under the assumption of LTE. 

The lines of the best matching spectral standard (GJ 908.1) are used as
input to represent the photospheric contribution except in the case of 
lithium.  For the $\lambda 6707$ resonance line, 
the low veiling observation of HJD 2450419.73527 (de-veiled) is used 
as input.  We proceed by computing the equivalent
widths of the (first) veiled and (then) extincted line profiles for the
highest veiling state we observe (v=6.1) over the above-mentioned grid of
slab temperatures and densities.  These equivalent widths are then de-veiled
in order to obtain consistency with the observed quantities plotted in 
Figures~\ref{allpts} to~\ref{mc}. 

Figure~\ref{slab} shows the results.  Plotted are the computed line
strengths as a function of hydrogen number density at a slab temperature
of 2000K, valid for the highest veiling observation (17 Jan 89; v=6.1).  
Horizontal lines show the observed (de-veiled) line strengths of each element.
The hatched area corresponds to the one-sigma uncertainty levels in these
observed quantities.  A slab temperature of 2000 K and density of $\log N_H 
\sim 16$ (marked by vertical line) is able to reproduce the observed line
strengths. In the case of MnI, we find that the slab provides no additional
line extinction, consistent with our observations.  In the case of RbI, 
the slab does furnish a small degree of extinction.  However, at this temperature
and at $\log N_H = 16$, the resulting line strength is less then 200 m\AA.  
Lines weaker than this will not be observable at our signal-to-noise in such a
highly veiled spectrum. 

We do not suggest that our computed slab parameters
are necessarily physically meaningful.  The accretion
flow is not well represented by an isothermal slab of uniform density
as has been shown by magnetic accretion models (\cite{m96,mch98,cg98}).
However, assuming the line enhancements are produced in approximately
the same region, such a simplified representation is useful and suggests
that a cold component of the accretion flow can yield the relative
line enhancements observed.

Observational constraints to magnetospheric accretion models are
commonly provided by the emission line spectrum as well as the excess
continuum emission.  Both diagnostics probe hot regions.
Muzerolle, Calvet, \& Hartmann (1998) find that average
temperatures of the accretion flow must reach 6000 - 10000 K in order
to reproduce the hydrogen emission lines. Martin (1996) computes self-consistent
models of the thermal structure of gas inflowing along magnetic field lines 
for a star with typical CTTS properties.  Temperatures reach peak values of
6500 - 8000 K, depending on the accretion rate.  However, the initial gas
temperature in the region where the magnetic field disrupts the disk
($3.4 - 5 R_*$, in his models) is arbitrarily fixed at 3000 K.  To date, we
lack an onservational parameter to constrain this quantity. Since we do not observe
large velocities of the excess absorption, this is precisely the region of
interest -- the region we believe the line enhancements are probing.
This is discussed further in \S 6.3.  

Investigating the effect of various accretion rates on the thermal structure, 
Martin finds that inefficient heating and efficient cooling at high accretion 
rates result in lower average temperatures in the accretion flow.  In 
particular, he considers 
accretion rates of $10^{-6}$, similar to that estimated for RW Aur 
(\cite{heg95}).  The thermal structure, though tied to 3000 K at the disk 
midplane, decreases to 2000 K before rising again as it approaches the 
stellar surface.  Such low temperatures are favorable for explaining the
line enhancements reported herein.

\begin{figure}
\caption{Modeled line equivalent widths are plotted as a function of hydrogen
number density (log scale) for a slab model of uniform temperature
(2000 K), of thickness $0.1 R_*$, covering 100\% of the stellar
surface.  Horizontal lines mark the observed equivalent widths 
in RW's highest veiling state (17 Jan 89).  At a density of 
$N_H \sim 10^{16}$, we recover consistent line strengths in all resonance
lines (to within the measured uncertainty levels shown as the hatched area).  
MnI shows no significant enhancement, consistent with observations.}
\label{slab}
\end{figure}

\subsection{Line Bisectors}

We search for any telltale asymmetries in the line profiles which may 
indicate bulk motion in the gas giving rise to the excess absorption.
For this purpose, we compute the line bisectors of the LiI and KI 
resonance lines obtained in November 1996 (see Table~\ref{obs}).  We 
concentrate on this subsample of observations due to the higher precision
of the wavelength calibration and radial velocity determinations.

Displayed in the upper panel of Figure~\ref{bisectors} are the line bisectors
of the LiI and KI resonance lines for two low veiling observations (HJD 2450419.73527,
HJD 2450419.84735 -- diamonds) and two high veiling observations (HJD 2450417.80447,
HJD 2450417.94248 -- asterisks).  The lowest veiling observation of RW Aur during 
the 1996 season is overplotted for reference.  A completely symmetric line profile 
would produce a perfectly vertical line bisector.  Blends produce systematic
deviations in the line bisectors of all observations, independent of veiling .  
But the high veiling observations show additional 
deviations towards the red.  This can be seen qualitatively in the middle
panels which contain the (de-veiled) profiles of the lowest (solid line;
HJD 2450419.73527; v=0.6) and the highest (dashed line; HJD 2450417.80447; v=2.7)  
veiling observations of November 1996.  It is evident that much of the excess
absorption generated by the accretion process is redshifted relative to line center.
Subtraction of the two veiling states yields redshifted residual absorption profiles
(lower panel) with line centroids at 9.8 km s$^{-1}$ for LiI and 19.5 km s$^{-1}$
for KI.  This lends further support to the hypothesis that the line
enhancements arise in an infalling gas and probe relatively low-velocity,
low-temperature regions near the co-rotation radius.

\begin{figure}
\caption{Line bisectors (upper panel) of the LiI and KI resonance lines are computed for two
low veiling (HJD 2450419.73527, HJD 2450419.84735) and two high veiling (HJD 2450417.80447,
HJD 2450417.94248) observations of RW Aur.  Diamonds mark the line 
bisectors of the low veiling observations while asterisks mark those of the high veiling
observations.  Deviations toward the red are clearly seen in the case of high veiling.
This can be seen qualitatively in the middle panel in which the lowest veiling (solid line;
HJD 2450419.73527; v=0.6) and the highest veiling (dashed line; HJD 2450417.80447; v=2.7) 
observation of each line is plotted (after de-veiling). Subtraction yields
residual absorption (lower panel; asterisks).  Line centroids (vertical line) of the 
residual absorption are computed via gaussian fits (solid line) and listed in the
bottom right corner.}
\label{bisectors}
\end{figure}

\subsection{Systematic Errors in Lithium Abundance Determinations}

Using two high-quality, high-spectral coverage observations of RW Aur
(HJD 2450417.80447,v=2.7; HJD 2450419.73527,v=0.6), we de-veil the spectral 
regions containing the lithium resonance lines at 6707.7 \AA~ and use the
resulting profiles to compute 
the LTE lithium abundances.  We employ the spectral synthesis code, SME, 
described in Valenti \& Piskunov (1996) and the line lists returned 
by the VALD database (\cite{pkrwj95}).  The code includes a Marquardt non-linear 
least squares algorithm which allows the user to simultaneously solve for selected 
stellar and/or atomic parameters which minimize the $\chi^2$ statistic.
In this manner, we obtain a lithium abundance of $\epsilon(N_{Li}) = 3.2$ for the low
veiling observation and $\epsilon(N_{Li}) = 4.4$ for the more highly veiled
observation.  The lower veiling observation gives us a better estimate
of the true lithium abundance on the stellar surface in light of the results
presented in \S 5.  We find a slightly depleted lithium abundance
relative to meteoritic values qualitatively in keeping with theoretical predictions.
However, at a veiling of 0.6, this is still likely to be an over-estimate.
As expected, the high veiling observation of RW Aur yields an abundance significantly
larger than meteoritic values as has previously been found for a handful of other CTTS 
(\cite{b91,m92}).  We conclude that the accretion-enhanced line strengths of 
lithium can lead to gross over-estimates in the true surface abundance.  

\section{Conclusions}

We present, in this communication, a study of high resolution echelle
spectra of the classical T Tauri star, RW Aur, taken sporadically over
a period of more than 10 years.  The continuum veiling distribution is
determined in the optical by comparing hundreds of stellar absorption lines
of a well-matching spectral template star with those of the program star.
Errors (both random and systematic) in the resulting veiling values are
investigated.  We find that systematic errors arising from improper
spectral classification are less than the actual line-to-line dispersion
in the veiling values and significantly smaller than
the magnitude of the veiling variations which RW Aur undergoes.

In the process of investigating LiI $\lambda 6707.7$ \AA~ as a potential 
veiling diagnostic (useful in cases of extremely high continuum veiling
which works to wash away all but the strongest absorption lines), we find 
that this particular absorption line is, in fact, enhanced by the accretion 
process.  Its line strength (after applying corrections for the effect of 
continuum veiling) more than doubles between the observation of lowest measured 
accretion and that of the highest measured accretion.  Searching for similar 
behavior among other resonance lines of low IP elements, we find that KI 
($\lambda 7699.9$ \AA) and TiI ($\lambda 4997.1$ \AA) are also enhanced by the 
accretion process, thereby eliminating the possibility that lithium undergoes 
true abundance variations as the disk supplies fresh lithium which convective 
mixing then works to deplete.  We conclude that the observed line enhancements 
are significant, and not merely a consequence of errors in veiling 
determinations, after finding no such enhancements in a large sample of 
absorption lines of different excitational potentials and of various IP 
elements. Approximately 83\% of the 29 lines studied show no correlation at 
all between their line strength and the respective continuum veiling.  

We find that the resonance line of the lowest IP element observed, KI (4.34 eV), 
presents the most sensitive correlation between veiling and line strength, 
followed by that of LiI (5.39 eV) and then TiI (6.82 eV).  The resonance line of MnI 
(7.43 eV) shows no similar behavior.  Finding it unlikely that cool magnetic surface
spots are responsible for the observed enhancements, we investigate the possibility
of an additional source of line opacity arising in a cool gas in the accretion 
flow.  Simple slab models are generated which successfully reproduce the relative 
line enhancements observed among the 4 resonance lines of KI, LiI, TiI, and MnI 
at a temperature of 2000 K and density of $\log N_H \sim 16$.  Models assume
a projected thickness of $0.1 R_*$ and a filling factor of 100\% (that is, the
entire stellar surface is assumed to be extincted by the accreting gas).
While providing line extinction, the slab does not introduce significant 
continuum opacity.

Line bisectors of the KI line profiles are computed, and comparison of those
of the low veiling observations with those of the high veiling observations
reveal that the excess absorption is preferentially redshifted relative to the
stellar rest-frame.  The bisectors of the low and high veiling states differ
by merely 10 km s$^{-1}$ for LiI and 20 km s$^{-1}$ for KI.  Centroids of
the residual absorption yield the same results.  This leads us to speculate 
that the line enhancements are indeed generated in an infalling gas located 
at the beginning of the accretion flow near the co-rotation radius.

As an exercise, the surface lithium abundances are computed using veiling-corrected 
$\lambda 6707.7$ \AA~ line profiles and ignoring the accretion-induced
enhancements.  We find that, as a consequence, surface lithium abundances
can be grossly over-estimated.  At a veiling of 0.6, we compute an LTE abundance
of $\epsilon (N_{Li})=3.2$, while at a veiling of 2.7, we compute an abundance 
of $\epsilon (N_{Li})=4.4$.  We hypothesize that the large abundances (often 
in excess of meteoritic values) reported for the highly accreting TTS 
suffer from such systematic errors. 

We conclude that the resonance lines of LiI, KI, and TiI are unreliable 
lines for measuring continuum veiling in classical TTS.  However, they
may act as an additional probe of the circumstellar environment which has
traditionally been left to the emission line spectrum.  We propose that a
cool component to the accretion column could explain the observed line
enhancements of the low IP resonance lines studied herein.  We also conclude
that surface lithium abundance measurements of classical TTS are 
subject to gross systematic errors.  Such errors can be 
minimized by studying systems, such as RW Aur, which present largely
varying accretion rates.  In its low accretion state, RW Aur presents
a slightly depleted surface lithium abundance, as is predicted by
theoretical models for a star of this age and mass.

\begin{acknowledgements}

This work was supported by the Funda\c c\~ao de Amparo \`a Pesquisa
do Estado do Rio de Janeiro (FAPERJ).  We thank E. Mart\'{\i}n for
his careful reading of this manuscript.

\end{acknowledgements}


\begin{thebibliography}{}

\bibitem[Anders \& Grevesse 1989]{ag89} Anders, E., Grevesse, N. 1989, 
Geochim. Cosmochim. Acta, 53, 197

\bibitem[Balachandran, Lambert, \& Stauffer 1988]{b88} Balachandran, S., 
Lambert, D.L., Stauffer, J.R. 1988, Astrophys. J., 333, 267

\bibitem[Basri \& Batalha 1990]{bb90} Basri, G.S., Batalha, C.C. 1990, 
Astrophys. J., 363, 654

\bibitem[Basri, Marcy, \& Graham 1996]{bmg96} Basri, G., Marcy, G.W., Graham, J.R.
1996, Astrophys. J., 458, 600

\bibitem[Basri \& Martin 1999]{bm99} Basri, G., Mart\'{\i}n, E.L. 1999, 
Astrophys. J., 510, 266

\bibitem[Basri, Mart\'{\i}n, \& Bertout 1991]{b91} Basri, G., Mart\'{\i}n, E.L., 
Bertout, C. 1991, Astron. Astrophys., 252, 625

\bibitem[Basri, Wilcots, \& Stout 1989]{bws89} Basri, G., Wilcots, E.,
Stout, N. 1989, Publ. Astron. Soc. Pac., 101, 528

\bibitem[Batalha et al. 1996]{bsbt96} Batalha, C.C., Stout-Batalha, N.M., 
Basri, G.S., Terra, M.A.O. 1996, Astrophys. J., Suppl. Ser., 103, 211

\bibitem[Bertout, Basri, \& Bouvier 1988]{bbb88} Bertout, C., Basri, G., 
Bouvier, J. 1988, Astrophys. J., 330, 350

\bibitem[Bodenheimer 1965]{b65} Bodenheimer, P. 1965, Astrophys. J., 142,
451

\bibitem[Bonsack \& Greenstein 1960]{bg60} Bonsack, W.K., Greenstein, 
J.L. 1960, Astrophys. J., 131, 83

\bibitem[Bouvier et al. 1993]{coy1} Bouvier, J., Cabrit, S., Fern\'andez, M.,
Mart\'{\i}n, E.L. and Matthews, J.M. 1993, Astron. Astrophys., 272, 176
 
\bibitem[Bouvier et al. 1995]{coy2} Bouvier, J., Covino, E., Kovo, O.,
Mart\'{\i}n, E.L., Matthews, J.M., Terranegra, L. and Beck, S.C. 1995,
Astron. Astrophys., 299, 89

\bibitem[Calvet \& Gullbring 1998]{cg98} Calvet, N. and Gullbring, E.
1998, Astrophys. J.. 509, 802
 
\bibitem[Cayrel et al. 1984]{c84} Cayrel, R., Cayrel de Strobel, G., 
Campbell, B., D\"appen, W. 1984, Astrophys. J., 283, 205

\bibitem[Choi \& Herbst 1996]{ch96} Choi, P.I. and Herbst, W. 1996,
Astron. J., 111, 283
 

\bibitem[Edwards et al. 1993]{e93} Edwards, S., Strom, S.E., Hartigan, P., 
Strom, K.M., Hilenbrand, L.A., Herbst, W., Attridge, J., Merril, K.M., 
Probst, R., Gatley, I. 1993, Astron. J., 106, 372

\bibitem[Finkenzeller \& Basri 1987]{fb87} Finkenzeller, U., Basri, G. 1987, 
Astrophys. J., 318, 823

\bibitem[Fischer 1998]{f98} Fischer, D.A. 1998, Ph.D. thesis, 
University of California, Santa Cruz

\bibitem[Garc\'{\i}a-Lopez, Rebolo, \& Mart\'{\i}n 1994]{g94} Garc\'{\i}a-Lopez, 
R.J., Rebolo, R., Mart\'{\i}n, E.L. 1994, Astron. Astrophys., 282, 518
 
\bibitem[Ghez, White, \& Simon 1997]{gws97} Ghez, A.M., White, R.J., 
Simon, M. 1997, Astrophys. J., 490, 353

\bibitem[Gliese \& Jahreiss 1991]{gliese} Gliese, W., Jahreiss, H. 1991, 
{\it Catalog of Nearby Stars}, Veroff. Astron. Rechen-Inst., Heidelberg, 
third edition

\bibitem[Gullbring et al. 1998]{g98} Gullbring, E., Hartmann, L., Briceno, C., 
Calvet, N. 1998, Astrophys. J., 492, 323

\bibitem[Hartigan, Edwards, \& Ghandour 1995]{heg95} Hartigan, P., Edwards, S., 
Ghandour, L. 1995, Astrophys. J., 452, 736

\bibitem[Hartmann et al. 1986]{hhsm86} Hartmann, L., Hewett, R., Stahler, S., 
Mathieu, R. 1986, Astrophys. J., 309, 275

\bibitem[Hessman \& Guenther 1997]{hg97} Hessman, F.V., Guenther, E.W. 1997, 
Astron. Astrophys., 321, 497

\bibitem[Horne 1986]{h86} Horne, K. 1986, Publ. Astron. Soc. Pac., 98, 609

\bibitem[James \& Jeffries 1997]{jj97} James, D.J., Jeffries, R.D. 1997, 
Mon. Not. R. Astron. Soc., 292, 252

\bibitem[Jeffries, James, \& Thurston 1998]{jjt98} Jeffries, R.D., 
James, D.J., Thurston, M.R. 1998, Mon. Not. R. Astron. Soc., 300, 550
 
\bibitem[Jones et al. 1997]{j97} Jones, B.F., Fischer, D., Shetrone, M., 
Soderblom, D.R. 1997, Astron. J., 114, 352

\bibitem[Jones, Fischer, \& Soderblom 1999]{j99} Jones, B.F., Fischer, D., 
Soderblom, D.R. 1999, Astron. J., 117, 330

\bibitem[Jones et al. 1996]{j96} Jones, B.F., Shetrone, M., Fischer, D., 
Soderblom, D.R. 1996, Astron. J., 112, 186

\bibitem[Keenan \& McNeil 1989]{km89} Keenan, P.C., McNeil, R.C. 1989, 
Astrophys. J., Suppl. Ser., 71, 245

\bibitem[Kraft 1967]{k67} Kraft, R.P. 1967, Astrophy. J., 150, 551

\bibitem[Magazz\`u, Rebolo, \& Pavlenko 1992]{m92} Magazz\`u, A., Rebolo, R., 
Pavlenko, Ya.V. 1992, Astrophys. J., 392, 159

\bibitem[Marsh 1989]{m89} Marsh, T. 1989, Publ. Astron. Soc. Pac., 100, 1032

\bibitem[Mart\'{\i}n \& Montes 1997]{mm97} Mart\'{\i}n, E.L., Montes, D. 1997, 
Astron. Astrophys., 318, 805

\bibitem[Mart\'{\i}n et al. 1994]{m94} Mart\'{\i}n, E.L., Rebolo, R., 
Magazz\`u, A., Pavlenko, Ya.V. 1994, Astron. Astrophys., 282, 503

\bibitem[Martin 1996]{m96} Martin, S.C. 1996, Astrophys. J., 470, 537

\bibitem[Mendes, D'Antona, \& Mazzitelli 1999]{m99} Mendes, L.T.S., 
D'Antona, F., Mazzitelli, I. 1999, Astron. Astrophys., 341, 174

\bibitem[Moore, Minnaert, \& Houtgast 1966]{solcat} Moore, C.E., Minnaert, M.G.J.,
Houtgast, J. 1966, The Solar Spectrum 2935 \AA~ to 8770 \AA, National Bureau of
Standards Monograph 61

\bibitem[Muzerolle, Calvet, \& Hartmann 1998]{mch98} Muzerolle, J., Calvet, N., 
Hartmann, L. 1998, Astrophys. J., 492, 743

\bibitem[Neuh{\"a}user et al. 1998]{n98} Neuh{\"a}user, R., Wolk, S.J.,
Torres, G., Preibisch, Th., Stout-Batalha, N.M., Hatzes, A.P., Frink, S.,
Covino, E., Walter, F.M., Alcala, J.M., Sterzik, M.F. and Wichmann, R.
1998, Astron. Astrophys., 334, 873

\bibitem[Nichiporuk \& Moore 1974]{nm74} Nichiporuk, W., Moore, C.B. 1974, 
Geochim. Cosmochim. Acta, 38, 1691

\bibitem[Patterer et al. 1993]{p93} Patterer, R.J., Ramsey, L., Huenemoerder, D.P., 
Welty, A.D. 1993, Astron. J., 105, 1519

\bibitem[Piskunov et al. 1995]{pkrwj95} Piskunov, N.E., Kupka, F., Ryabchikova,
T.A., Weiss, W.W. and Jeffery, C.S. 1995, Astron. Astrophys., Suppl. Ser.,
112, 525

\bibitem[Randich et al. 1996]{r96} Randich, S., Aharpour, N., Pallavicini, R., 
Prosser, C.E., Stauffer, J.R., Schmitt, J.H.M.M. 1996, Proceedings of the
9th Cambridge Workshop on Cool Stars, Stellar Systems, and the Sun, Astronomical
Society of the Pacific Conference Series, volume 109, edited by Roberto Pallavicini
and Andrea K. Dupree, p.379

\bibitem[Randich et al. 1998]{r98} Randich, S., Mart\'{\i}n, E.L., 
Garc\'{\i}a L\'opez, R.J., Pallavicini, R. 1998, Astron. Astrophys., 
333, 591

\bibitem[Rydgren \& Vrba 1983]{rv83} Rydgren, A.E. and Vrba, F.J. 1983,
Astrophys. J., 267, 191

\bibitem[Soderblom et al. 1993a]{s93b} Soderblom, D.R., Fedele, S.B., 
Jones, B.F., Stauffer, J.R., Prosser, C.F. 1993a, Astron. J., 106, 1080

\bibitem[Soderblom et al. 1993b]{s93a} Soderblom, D.R., Jones, B.F., 
Balachandran, S., Stauffer, J.R., Duncan, D.K., Fedele, S.B., Hudon, J.D.
1993b, Astron. J., 106, 1059

\bibitem[Soderblom et al. 1995]{s95} Soderblom, D.R., Jones, B.F., 
Stauffer, J.R., Chaboyer, B. 1995, Astron. J., 110, 729

\bibitem[Stauffer et al. 1998]{s98} Stauffer, J.R., Schultz. G., 
Kirkpatrick, J.D. 1998, Astrophys. J., Lett., 499, 199

\bibitem[Stout-Batalha 1997]{mythesis} Stout-Batalha, N.M., Ph.D. thesis,
University of California, Santa Cruz

\bibitem[Stout-Batalha \& Vogt 1999]{sv99} Stout-Batalha, N.M., 
Vogt, S.S. 1999, Astrophys. J., Suppl. Ser., 123, ???

\bibitem[Thorburn et al. 1993]{t93} Thorburn, J.A., Hobbs, L.M., Deliyannis, C.P., 
Constantine, P., Pinsonneault, M.H. 1993, Atrophys. J., 415, 150

\bibitem[Tonry \& Davis 1979]{td79} Tonry, J. and Davis, M. 1979, Astron. J.,
84, 1511

\bibitem[Valenti et al. 1993]{v93} Valenti, J.A., Basri, G., Johns, C.M. 1993, 
Astron. J., 106, 2024

\bibitem[Valenti \& Piskunov 1996]{vp96} Valenti, J.A. and Piskunov, N. 1996
Astron. Astrophys., Suppl. Ser., 118, 595

\bibitem[Ventura et al. 1998]{v98} Ventura, P., Zeppieri, A., Mazzitelli, I., 
D'Antona, F. 1998, Astron. Astrophys., 331, 1011

\bibitem[Vogt 1987]{v87} Vogt, S.S. 1987, Publ. Astron. Soc. Pac., 99, 1214

\bibitem[Zappala 1972]{z72} Zappala, R.R. 1972, Astrophys. J., 172, 57

\end{thebibliography}
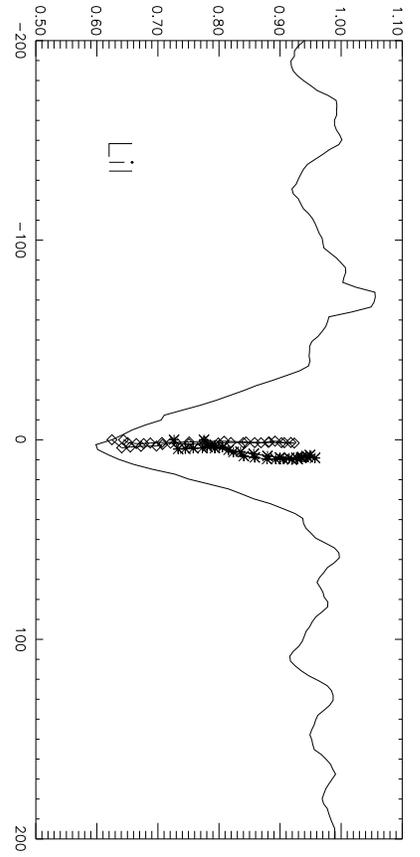
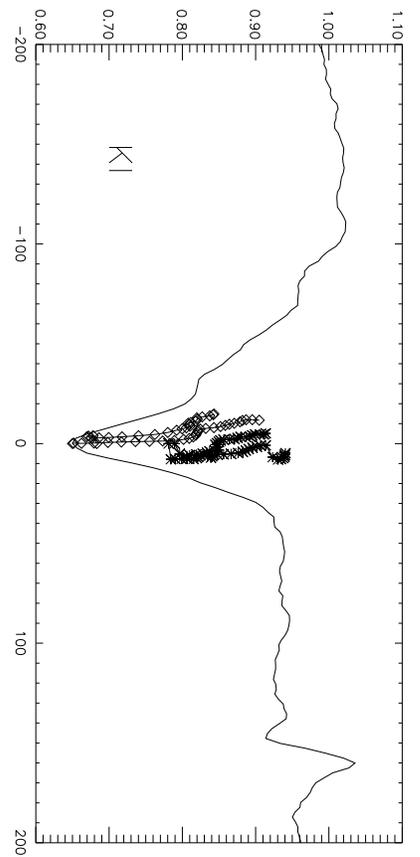
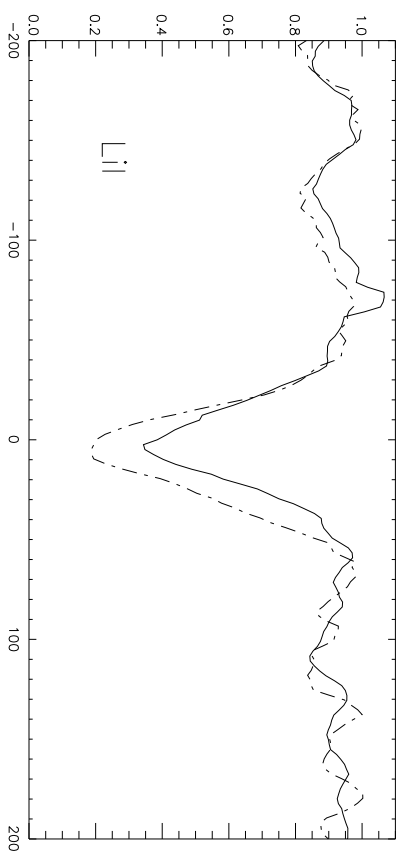
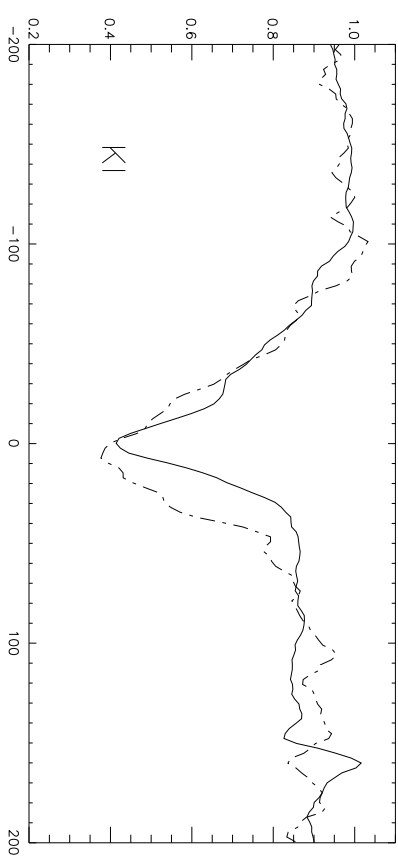
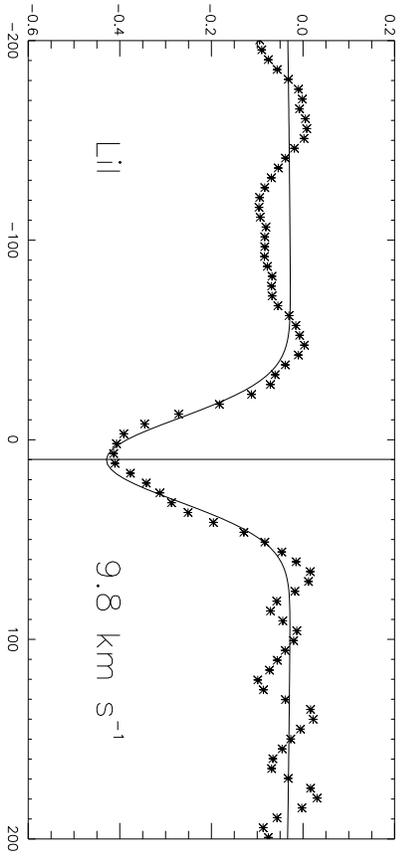
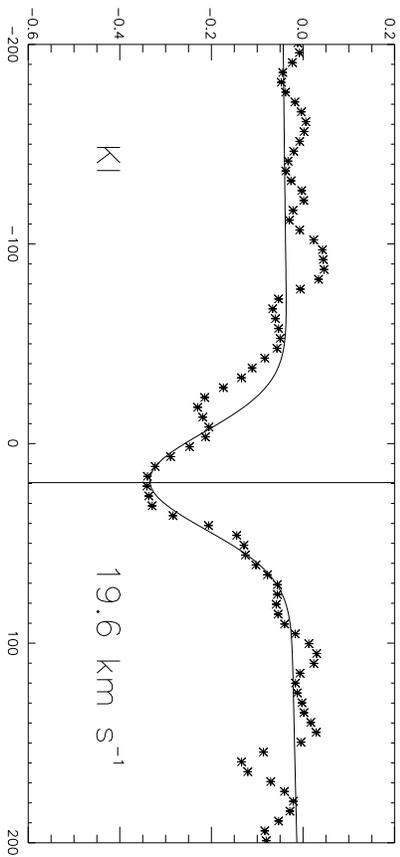
\end{document}